\documentstyle[12pt,mymoriond]{article}
\begin{document}
\vspace*{-10mm}
\noindent
To appear in {\it Dwarf Galaxies and Cosmology}, the Proceedings of the 
XVIII$^{th}$ Moriond Astrophysics Meeting, Les Arc, France, 1998, 
eds. T. X. Thu\^{a}n, C. Balkowski, V. Cayatte, \& J. Tr\^{a}n Thanh V\^{a}n 
(Editions Fronti\`{e}res, Gif-sur-Yvette)\\[-10mm]

\heading{Globular Clusters in Dwarf Elliptical Galaxies}

\author{B. W. Miller$^{1}$, H. C. Ferguson$^{2}$, J. Lotz$^{3}$, 
M. Stiavelli$^{2}$, B. C. Whitmore$^{2}$} 
{$^{1}$ Leiden Observatory, Leiden, NL.}  
{$^{2}$ Space Telescope Science Institute, Baltimore, USA.}
{$^{3}$ Johns Hopkins University, Baltimore, USA.}

\begin{moriondabstract}
Globular clusters (GCs) provide one way of determining when the most
important periods of star formation in dwarf elliptical (dE) galaxies
occurred.  Thus, they are also useful for comparing the star formation
histories of dwarf and giant galaxies.  We present results on the GC
systems of 24 nearby dE galaxies  
imaged with the {\it Hubble Space Telescope}.  We find that the
specific GC frequency, $S_N$, increases with $M_V$.  Also, $S_N$ in
dEs is more like the values found in giant ellipticals rather than in
spirals, and the mean $S_N$ for nucleated dEs is roughly a factor of 2
higher than for non-nucleated dEs.  The colors suggest that most of the GCs
formed before the bulk of the stars in the galaxies.
\end{moriondabstract}

\section{Introduction}

Dwarf elliptical (dE) galaxies make up most of the objects in the
steeply rising, low-luminosity end of the galaxy luminosity function
in galaxy clusters \cite{t98}\cite{jpsp}.  Besides being the most
numerous type of galaxy in the universe, in hierarchical clustering
scenarios of galaxy formation progenitors of dEs are the building
blocks of massive galaxies.  Therefore, solutions to the questions of
when dEs formed and how they evolve have important cosmological
implications.  Unfortunately, there is still little consensus on these
issues or even about the connections between different types of dwarf
galaxies (see \cite{fb94} for a review).

The globular cluster systems of dE galaxies may provide an observational
probe of their most important episodes of star formation.  Globular
clusters (GCs) are thought to form under the extreme pressures produced by
very vigorous star formation \cite{ee97}\cite{hhm98}.  Since GCs have
instantaneous burst stellar populations, their integrated light contains
information about their ages and metallicities. Also, the specific globular
cluster frequency ($S_N = N_{c}10^{0.4(M_V+15)}$) 
is dependent on Hubble type; it is a factor of 2--4
higher for giant ellipticals than for spirals and irregulars \cite{h91}.
This suggests a test for whether dEs have had star formation histories more
like giant ellipticals or to irregulars.  If they are related to Es, then
they should have $S_N \sim 2-5$, if they are more like late-type disks then
one would expect $S_N\sim1$.

To test this hypothesis and study the properties of the GCs and nuclei of
dE galaxies, we have begun a {\it Hubble Space Telescope (HST)} snapshot
survey of dEs in nearby groups and clusters.  The original sample was split
evenly between nucleated (dE,N) and non-nucleated (dE,noN) galaxies.  
In {\it HST} Cycle 6 we
obtained observations of 24 galaxies in the Virgo and Fornax Clusters and
the Leo Group that are suitable for measuring $S_N$.  We are able to measure 
($V\!-\!I$) colors of GCs down to $V\approx25$ \cite{de6}.

\section{Results}

The main results of this survey at present are summarized below.

\begin{itemize}

\item {\bf Nuclei:} The high resolution of {\it HST} allows us to
refine the classification of the dEs into nucleated and non-nucleated
categories.  For example, FCC~324 was classified as a non-nucleated dS0
from ground-based photographic material, but now clearly shows a
nucleus.  VCC~9 had been classified as nucleated but the ``nucleus''
is offset from the centroid of the background light by about 160~pc.
Therefore, we classify it as non-nucleated.  Three other galaxies do
not meet our criteria for being nucleated but their more luminous GCs
are within $\sim60$~pc of the galaxies' centroids.  These latter four
galaxies may have {\it offset nuclei} and they are shown in
Figure~1.

\begin{figure}
\epsscale{0.58}
\plotone{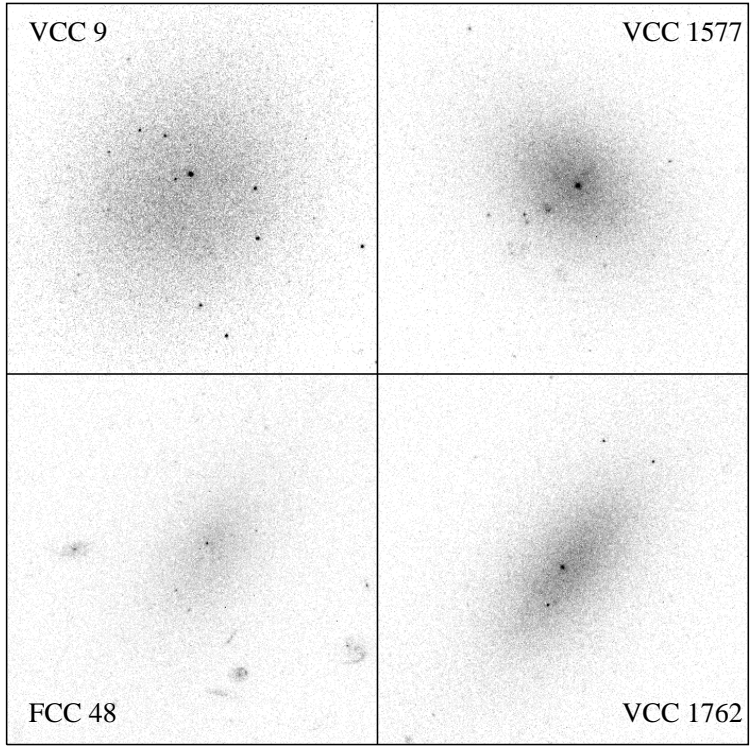}
\caption{{\it HST} snapshots of four dE galaxies with possible offset
nuclei.  Each image shows a physical region about 3~kpc across.}
\end{figure}

\item {\bf Luminosity Function:} The luminosity function (LF) of the
GC candidates from the entire sample is consistent with the standard
Gaussian-shaped GCLF with a peak at $M_V^0=-7.4$ and a width
$\sigma_V\approx1$~mag.  Also, with a mean color of $\langle
V\!-\!I\rangle\approx0.95$, most of the GCs appear to be
old and metal-poor ($[{\rm Fe/H}]\approx-1.5$) like GCs in the Galaxy
and in nearby giant ellipticals.

\item {\bf Metallicities:} If differences in $(V\!-\!I)$ are due to 
metallicity, then in general we find
\begin{displaymath}
[{\rm Fe/H}]_{GC} <  [{\rm Fe/H}]_{nuc} <  [{\rm Fe/H}]_{gal}.
\end{displaymath}
The GCs are between 0.0 and 0.4 dex more {\it metal-poor} than the
stars in bodies of the galaxies.  The GCs in a couple dEs have large
color spreads that imply metallicity spreads of up to 1 dex.  The most
luminous nuclei are $\sim0.3$ dex more {\it metal-rich} than the GCs
in these galaxies.

\item {\bf Trends in $S_N$:} There is a trend of increasing $S_N$ with
$M_V$ (Figure~2). Also, there is evidence for a difference between
dE,N and dE,noN galaxies.  The dE,Ns define the upper envelope of this
trend (solid curve).  The trend for the dE,noN is much shallower, it
is basically flat (dashed curve).  The galaxies with offset nuclei
shown in Figure~1 seem to have $S_N$ more like the nucleated galaxies.

\begin{figure}
\plotfiddle{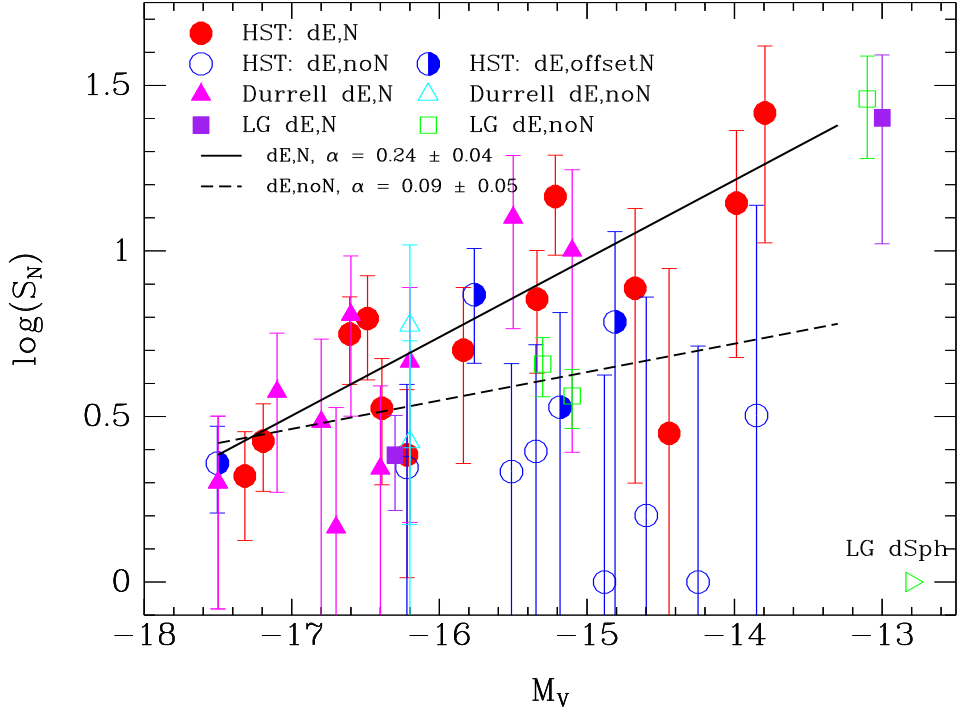}{8.7cm}{-90}{52}{52}{-210}{285}
\caption{$\log(S_N)$ vs. $M_V$ for all dE galaxies with measured $S_N$.
For galaxies with $S_N=0$, $\log(S_N)$ has been set to 0.0.
The circular symbols are from the {\it HST} snapshot sample \cite{de6}.
Triangles are from the ground-based work of Durrell et al. \cite{d96}
and squares are galaxies in the Local Group (LG).  The triangle labeled
``LG dSph'' represents the $\sim12$ LG dwarf spheroidals (non-nucleated)
that have $S_N=0$. The lines are weighted least-squares fits.}
\end{figure}

\item {\bf Averages for $S_N$:} The differences in $S_N$ suggested by
Figure~2 are also reflected in the average values of $S_N$ for the two types
are given below.  These averages include dE galaxies in the Local Group and
Durrell et al.'s \cite{d96} ground-based sample with $M_V<-13.5$.  The
uncertainties are standard deviations of the mean.

\begin{center}
\begin{tabular}{lc} 
Type & $\langle S_N\rangle$ \\ \hline         
dE,N & $6.4\pm1.2$ \\
dE,noN & $3.3\pm0.5$ \\ \hline
\end{tabular}
\end{center}

\end{itemize}

\section{Discussion}

The basic result of this work is that 
the mean values of $S_N$ in dEs is in the range 3--6, very similar to
values for $S_N$ in giant ellipticals in rich clusters \cite{h91}.  Thus it
would seem that most dEs, and especially the dE,Ns, have experienced similar
star formation histories as giant Es.  The colors of the GCs suggest that 
most of the 
cluster formation occurred very early out of relatively metal-poor gas.
Subsequent star formation from more enriched gas produced most of the
stars in the bodies of the galaxies.

The differences in $S_N$ between dE,Ns and dE,noNs may be a result of 
environment.  Nucleated dEs are preferentially found in the centers of 
rich clusters and have spatial distributions like giant Es \cite{fs89}.
Thus, they would have formed in a high pressure, turbulent environment
that would be conducive to GC formation \cite{ee97}.  Gravitational torques
from interactions could drive gas towards
the centers and help with the formation of nuclei.
On the other hand, dE,noNs are more uniformly distributed, like the
spirals and irregulars \cite{fs89}.  Thus, they may have formed in 
lower-pressure environments less suited to the formation of GCs or nuclei.
Numerical simulations would be useful for testing this scenario.
Finally, some dE,noNs have $S_N=0$, making themselves candidates for
being stripped irregulars.

The trend of $S_N$ with $M_V$ is more difficult to explain.  One possibility
is that dynamical friction is causing many of the more massive GCs to 
conglomerate into the nucleus.  The dynamical friction timescales are short
enough for this to occur.  A second
option is that the most luminous dEs are stripped spirals; the bulges
of spirals do have similarities to dE,Ns \cite{c97}.

This work has shown the usefulness of using GCs to compare the star
formation histories of dwarf galaxies and giant galaxies.  However,
several issues still need to be resolved.  First, it is not clear whether the
trends in $S_N$ continue to fainter magnitudes or whether there
is just an increase in scatter.  Therefore, in {\it HST} Cycle~7 we
will image up to 30 more faint dEs.  Also, the mean value and scatter
in $S_N$ for dwarf irregulars (dIs) is very uncertain since very few
dIs have $S_N$ measurements.  Therefore, we will also image up to 30
dIs in Cycle~7.  Hopefully, this will provide a clearer picture of
the connection between different types of dwarf galaxies.

\acknowledgements

B.W.M. would like to thank the Leids Kerkhoven Bosscha Fonds and the
Netherlands Foundation for Research in Astronomy (ASTRON) for 
financial support to attend this meeting.


%
\vfill
\end{document}